\documentstyle[twoside]{article}

\newcount\cislo
\newcount\prvastrana
\newcount\poslednastrana
\def\title#1{\begin{center}{\bf #1}\end{center}\smallskip}
\def\author#1#2{\begin{center}{\bf #1}\\ {\sl #2}\end{center}}

\def\abstract#1{\begin{quotation}\noindent{\small#1}\end{quotation}\medskip}

\def\caption#1{\noindent\small#1}

\def\ps@headings{\let\@mkboth\markboth

\def\@oddfoot{\ifnum\thepage=\prvastrana{
\noindent{\small \publ}\rm\thepage}
\else{}\fi}

\def\@evenfoot{}

\def\@evenhead{\ifnum\thepage=10000{\thepage\quad\sl{\runauthor: \nazov\hfill}}
\else{\thepage\hfil{\sl\runauthor}\qquad}\fi}

\def\@oddhead{\ifnum\thepage=\prvastrana{}
\else \qquad{\sl\shorttitle}\hfill\thepage\fi}

\def\sectionmark##1{\markboth {\uppercase{\ifnum \c@secnumdepth
>\z@
 \thesection\hskip 1em\relax \fi ##1}}{}}\def\subsectionmark##1{\markright
{\ifnum \c@secnumdepth >\@ne
 \thesubsection\hskip 1em\relax \fi ##1}}}

\def\refer#1#2#3#4#5{#1:\ {\sl #2}\ {\bf #3}\ {(#4)}\ #5;\ }

\def\runauthor{Martina Brisudov\'a}
\def\shorttitle{Bound state studies in light-front QCD of mesons ...}

\begin{document}
\title{Bound state studies  in light-front QCD
of mesons containing at least one heavy 
quark.}
\author{Martina Brisudov\'a }{Department of
Physics, The Ohio State University, Columbus, OH 43210}

\abstract{We present the first numerical QCD bound-state calculation based 
on a 
renormalization
group-improved light-front Hamiltonian formalism. The QCD Hamiltonian is
determined to second order in the coupling, and it includes two-body 
confining interactions. We make a momentum expansion, obtaining an equal-
time-like 
Schr{\H{o}}dinger equation. This is solved for quark-antiquark constituent
states, and we obtain a set of self-consistent parameters by fitting B 
mesons. Applying the approach to charmonium spectra leads to a prediction 
of the hyperfine splitting between J/$\psi$(1S) and $\eta_c$(1S) which is 
in good agreement with experiment.}
%
%
\setcounter{enumi}{1}

\section{Introduction }
Recently,  a new approach to QCD
bound states has been proposed \cite{1,2}.
The goal
 is to build a bridge between QCD and a constituent quark model (CQM).
It has been
argued that it is convenient to use a light-front formulation of the theory,
because on the light-front it is possible to make the vacuum trivial simply
by  implementing a small longitudinal cutoff. As a result,
 all partons in a hadronic state
are connected to the hadron, instead of being disconnected excitations in a
complicated medium. The price to pay is a considerably more complicated
renormalization problem.

The new approach consists of two (major) steps. The first step is to find an
effective Hamiltonian, starting from the canonical light-front Hamiltonian 
regulated by a large cutoff (for details on the cutoff scheme we refer the
reader to refs. \cite{1,2}). Unitary transformations are used to find 
counterterms which remove the dependence on the large cutoff, and to bring 
the Hamiltonian  towards band-diagonal form with respect to free
light-front energies. These transformations  form a
renormalization (semi)group. This is repeated until the Hamiltonian is
 band-diagonal 
with respect to a typical
hadronic scale. At the end of this first step the
Hamiltonian is still a complicated field theory Hamiltonian, but it does not
couple states which differ in their free light-front energies by more than the
hadronic scale. Instead, it contains effective potentials. It was shown that
the effective potentials contain a Coulomb and a logarithmic confining
potential already at order $g^2$ \cite{2}. 

In the second step
we want to solve this Hamiltonian and find its spectrum. We divide the
effective Hamiltonian into a part $H_0$ which is solved nonperturbatively, and
 a part $V$
which is then calculated in bound-state perturbation theory.
We want 
to choose $H_0$ so that it is manageable (i.e.
something that we can solve) and 
we want it to 
contain the essential physics. Taking hints
from the constituent quark model,  
we include two body potentials and use constituent masses in the
$H_0$, but do not include emission and absorption. Any approximations can
always be done by adding a term to $H_0$ and subtracting it from $V$
which is treated in bound state perturbation theory. Not including the emission
and absorption in the $H_0$ has an important consequence: different Fock
states decouple. We can thus solve few body problems. The consistency of 
this procedure has to be checked in
bound state perturbation theory.

In this talk 
I present one of the simplest QCD bound state calculation 
based on 
similarity transformations \cite{3}  and using 
coupling coherence \cite{4}.
We find the
effective Hamiltonian to order $g^2$, 
 and then solve for $q{\bar{q}}$ bound
states. 
The calculation is carried
through for
quarks of arbitrary but nonzero masses. At the end, we concentrate on mesons
containing at least one heavy quark. 

\section{The effective Hamiltonian to order $g^2$}
The effective Hamiltonian, 
which generated by the similarity transformation to 
order $g^2$,
is band-diagonal in light-front energy with respect 
to a hadronic scale 
${\Lambda^2\over{{\cal P}^+}}$,   and it can be written as:
\begin{eqnarray}
H_{\rm eff} = H_{\rm free} + V_1 +  V_{2} +  V_{2 \ {\rm eff}} \  \  ,
\end{eqnarray}
where $H_{\rm free}$ is the light-front kinetic energy (we remind the 
reader that the light-front kinetic energy of a particle with transverse 
momentum $\vec{p}^{\perp}$ and longitudinal momentum $p^+$ is 
${{p^{\perp}}^2 + m^2 \over{p^+}}$), $V_1$ is ${\cal O}(g)$
emission and absorption with
nonzero matrix elements
only between states with energy difference smaller than the
hadronic scale ${\Lambda^2\over{{\cal P}^+}}$, $V_{2} $ is
${\cal O}(g^2)$
instantaneous interaction, and $V_{2 \ {\rm eff}}$ includes
the effective
interactions generated by similarity, also ${\cal O}(g^2)$.
The effective interactions generated to this order contain one-body and
 two-body operators.
In particular, the effective one-body operator is:
\begin{eqnarray}
{\alpha_{\Lambda} C_F \over{2\pi P^+}} \left\{
2 {P^+\over{{\cal P}^+}}\Lambda^2 \log\left({P^+\over{\epsilon {\cal P}^+}}
\right)
+ 2 {P^+\over{{\cal P}^+}}\Lambda^2
\log {x_a^2 {P^+\over{{\cal P}^+}}\Lambda^2 \over{x{P^+\over{{\cal P}^+}}
\Lambda^2 +m_a^2  }} \right. \nonumber\\
- {3\over{2}}{P^+\over{{\cal P}^+}}\Lambda^2
+{1\over{2}}{m_a^2 {P^+\over{{\cal P}^+}}\Lambda^2
\over{x_a{P^+\over{{\cal P}^+}}\Lambda^2 +m_a^2  }} \nonumber\\
 \left.
+ 3 {m_a^2\over{x_a}}\log
 {m_a^2\over{x_a {P^+\over{{\cal P}^+}} \Lambda^2 +m_a^2  }} \right\} \  \  ,
\end{eqnarray}
where $x_a= {p_a^+\over{P^+}}$ 
is the longitudinal fraction of the momentum carried by the 
constituent under consideration, $m_a$ is its mass, $P^+$ is the total 
longitudinal momentum of the state, ${\cal P}^+$ is the longitudinal scale 
required in the cutoff by dimensional arguments, and $\epsilon$ is an 
infrared cutoff which is to be taken to zero. The divergence in the 
effective one-body operator exactly cancels against the divergence in the 
effective two-body operator if the state is a color singlet.

The effective two-body operators have the following matrix elements between 
states containing a quark of momentum $\vec{p}_i$ and an antiquark of momentum 
$\vec{k}_i$, $i=1,2$ referring to the initial and final state, respectively:
\begin{eqnarray}
  & -g_{\Lambda }^2 \bar{u}(p_2, \sigma _2) \gamma ^{\mu} u(p_1, \sigma _1)
\bar{v}(k_2, \lambda_2) \gamma ^{\nu} v(k_1, \lambda _1)
\langle T_a T_b \rangle
\nonumber\\
 & \times \left[
 {1\over{q^+}}
D_{\mu \nu}(q)
\left({\theta(\vert D_1\vert-{\Lambda^2 \over{{\cal P}^+}})
\theta(\vert D_1\vert -\vert D_2\vert )\over{D_1}}
+ {\theta(\vert D_2\vert-{\Lambda^2 \over{{\cal P}^+}})
\theta(\vert D_2\vert -\vert D_1\vert )\over{D_2}}\right)
\right] 
\end{eqnarray}
where $\sigma _i$ and $\lambda_i$ are  light-front helicities of the quark and 
antiquark, respectively,
$u(p, \sigma )$ and $v(k, \lambda )$ are their spinors,
$D_{\mu \nu}(q) = {{q^{\perp}}^2\over{{q^+}^2}}\eta_{\mu}\eta_{\nu}
 + {1\over{q^+}}
\left(\eta_{\mu}{q^{\perp}}_{\nu} + \eta_{\nu}{q^{\perp}}_{\mu}\right)
- g^{\perp}_{\mu \nu}$ is the gluon propagator in light-front gauge,
$\eta _\mu = ( 0, \eta _+ = 1,0,0)$,
$\vec{q} = \vec{p}_1 - \vec{p}_2$ is the exchanged momentum and
$q^- ={ {q^{\perp}}^2\over{q^+}}$. $D_1$, $D_2$ are energy denominators:
$D_1 = p_1^- -p_2^- -q^-$ and $D_2 = k_2^- -k_1^- - q^-$.  

\section{Bound state calculation}
For the purpose of the bound-state calculation we  divide the effective 
Hamiltonian into two parts: $H_0$ which is solved nonperturbatively, and 
$V\equiv H_{eff} - H_0$ which is solved in bound-state perturbation theory. 
In $H_0$, we include nonrelativistic limit of the kinetic 
energy, the self-energies, and the 
rotationally symmetric part of the $\eta_{\mu} \eta_{\nu}$ term of the two-body 
interactions (both instantaneous and  generated by similarity) \cite{5}.
Let the masses of the constituents be $m_a$ and $m_b$, and 
\begin{eqnarray}
M_{ab} \equiv m_a +m_b .
\end{eqnarray}
In the nonrelativistic limit, the light-front scale 
${\Lambda^2\over{{\cal P}^+}}$ is naturally replaced by 
${\cal L}\equiv{\Lambda^2\over{{\cal P}^+}}{P^+\over{M_{ab}}}$,
which carries dimension of mass \cite{5}.

The Hamiltonian $H_0$ is:
\begin{eqnarray}
H_0 = 2M_{ab}\left[ -{1\over{2m}}\vec{\nabla}^2 +\tilde{\Sigma}
- { C_F \alpha \over{r}} +{ C_F \alpha {\cal L} \over{\pi}} V_0({\cal L}r)
\right],
\end{eqnarray}
\noindent where $m$ is the reduced mass and 
\begin{eqnarray}
V_0({\cal L}r)   & = & 
2 \log {\cal R}-2 Ci({\cal R}) +4 {Si({\cal R})\over{{\cal R}}}
-2 {(1-\cos{\cal R})\over{{\cal R}^2}}+2{\sin{\cal R}\over{{\cal R}}}
-5 +2\gamma  \  \  , 
\end{eqnarray}
where $\gamma$ is Euler constant.  
 $\tilde{\Sigma }$ contains the finite shift produced by the self-energies 
after subtracting terms needed to make the confining potential vanish at 
the origin:
\begin{eqnarray}
\tilde{\Sigma } &  =  & 
 {\alpha C_F  {\cal L}\over{ 2 \pi}} 
\left[  
\left( 1+{3m_a\over{4{\cal L}}}\right)
 \log \left( {m_a\over{ {\cal L}+m_a}} \right)  + 
\left( 1+{3m_b\over{4{\cal L}}}\right)
\log \left( {m_b\over{ {\cal L}+m_b}}\right) \right.
\nonumber\\
  & \ \ \   &  \  \  \  \  \  \  \  \  \  \   \left.
+{1\over{4}}{m_a\over{ {\cal L}+m_a}}+{1\over{4}}{m_b\over{ {\cal L}+m_b}}
+{5\over{2}} 
\right]  .
\end{eqnarray}

\subsection{Schr{\H{o}}dinger equation.}

We now want to find the mass of a $q\bar{q}$ bound state and its wave function 
$\psi (\kappa^{\perp}, x)$:
\begin{eqnarray}
\vert P \rangle = \int {d^2\kappa^{\perp} \ dx \over{2(2\pi)^3 
\sqrt{x(1-x)}}}\psi(\kappa^{\perp},x) 
b^{\dagger} d^{\dagger} \vert 0 \rangle \  \  .
\end{eqnarray}
We use  a Lorentz-invariant normalization for the states:
$$\langle P' \vert P \rangle = 2(2\pi)^3 P^+ \delta ^3(\vec{P}-\vec{P'}),$$
and the wave function is normalized to one:
$$  \int {d^2\kappa^{\perp} \ dx \over{2(2\pi)^3 
}}\vert \psi(\kappa^{\perp},x) \vert^2 = 1 .$$
The bound state satisfies:
\begin{eqnarray}
H_{0} \vert P \rangle = {\cal M}^2 \vert P \rangle \  \  , 
\end{eqnarray}
where ${\cal M}^2$ is the invariant mass of the bound state.
Let the mass of the bound state be
\begin{eqnarray}
{\cal M}^2 = (m_a +m_b)^2 +2(m_a+m_b) E \   \  ,
\end{eqnarray}
which defines $E$. 

After substituting for  $H_0$ and ${\cal M}^2$ in 
equation $(9)$,  some straightforward algebra leads to a bound-state equation 
for the wave function $\psi$:
\begin{eqnarray}
 M_{ab} \left(E - \tilde{\Sigma}
 + {1\over{2m}}{d^2\over{d\vec{r}^2}}\right) \psi (\vec{r}) =
 M_{ab} \left[{-\alpha C_F \over{r}}+{\alpha C_F {\cal L}\over{\pi}}
V_{\rm conf}({\cal L}r)
\right]\psi (\vec{r}) \  \  .
\end{eqnarray}
It is convenient to use a dimensionless separation ${\cal R}={\cal L}r$ that 
naturally arises in the confining piece of the potential, and to absorb 
$-\tilde{\Sigma}$
into a definition of the 
eigenvalue $\tilde{E}$ of the Schr{\H{o}}dinger equation. When extracting the
bound state mass, $-\tilde{\Sigma}$  has to be subtracted.
The bound-state equation in  dimensionless form is:
\begin{eqnarray}
\left[ -{{\cal L}^2\over{2m}}{d^2\over{d\vec{{\cal R}}^2}}
+ {\cal L} \alpha C_F \left( {1\over{\pi}}V_{\rm conf}({\cal R}) + V_{\rm coul}(
{\cal R})\right) \right] \psi({\vec{{\cal R}}}) =
 \tilde{E}\psi({\vec{{\cal R}}}) \  \  .
\end{eqnarray}
Multiplying both sides of the equation by $2m/{\cal L}^2$ 
and introducing a dimensionless coupling and eigenvalue:
\begin{eqnarray}
c & \equiv & {2m\alpha C_F \over{{\cal L}}} , 
\\
e & \equiv & {2m \tilde{E} \over{{\cal L}^2}} ,
\end{eqnarray}
we obtain a Schr{\H{o}}dinger equation, which depends only on dimensionless 
variables:
\begin{eqnarray}
\left[ -{d^2\over{d\vec{{\cal R}}^2}}
+ c \left(
 {1\over{\pi}}V_{\rm conf}(\vec{\cal R}) + V_{\rm coul}(
{\cal R})\right) \right] \psi({\vec{{\cal R}}}) =
 e \psi({\vec{{\cal R}}}) \  \  .
\end{eqnarray}

This form is advantageous for numerical study, but moreover, it is 
quite general
- one obtains an equation of this form for any quark-antiquark 
systems and any choice of the confining potential 
in the nonrelativistic limit, regardless of the masses, providing 
they are nonzero.  
For different systems ${\cal L}$, $c$, $e$ would differ,
 but the resulting dimensionless 
Schr{\H{o}}dinger
equation will be the same. Thus in the leading order,  {\it qualitative}
characteristics of spectra depend only on one particular combination of the 
masses
and the coupling, as seen from equations $(13)$ and $(14)$ \cite{5}.

\subsection{Results and conclusion}

We refer the reader to ref. \cite{5} for more details on what one can learn 
from this dimensionless Schr{\H{o}}dinger
equation, and the dimensionless results. 
We applied the approach to B mesons \cite{5}, 
and we obtained a set of self-consistent parameters.
Applying the approach to charmonium, we used 1S, 1P and 2S levels to fit 
our parameters ($m_c =1.5$ GeV, $\Lambda = 1.7$ GeV, $\alpha =0.5$) and 
then predicted the hyperfine splitting in the charmonium ground state 
(the splitting between J/$\psi$(1S) and $\eta _c$(1S) )  to be about $0.13$ GeV
\cite{6}.

Our study shows that the new approach proposed by Wilson et al. is  
promising.

\section*{Acknowledgements}
The work I presented has been done in collaboration with Robert Perry and 
with involvement of Ken Wilson.
I would like to thank the organizers for giving me the opportunity to present 
my work in front of my home crowd. I would like to thank Palo Stri\v zenec 
and Palo {\v S}{\v t}avina for useful discussions and their warm hospitality.

\end{document}